\documentclass[12pt]{iopartmod1}
\usepackage{color, graphicx}
\usepackage{bm}

\usepackage{iopams}

\font\smalltesti=cmmi8 at 8.0pt
\font\smallseveni=cmmi7 at 5.83pt
\font\hscript=eufb10 scaled \magstep 1  
\def\hsp{{\hscript h}}

\begin{document}

\title[Lower-order ODEs for twisting type N Einstein spaces via CR geometry]
{Lower-order ODEs to determine new twisting \\ type N Einstein spaces via CR geometry}

\author{Xuefeng Zhang and Daniel Finley}

\address{Department of Physics and Astronomy, University of New Mexico, Albuquerque, NM 87131 USA}
\eads{\mailto{zxf@unm.edu} and \mailto{finley@phys.unm.edu}}

\begin{abstract}
In the search for vacuum solutions, with or without a cosmological constant, $\Lambda$,
of the Einstein field equations of Petrov type N with twisting principal null directions,
the CR structures to describe the parameter space for a congruence of such null vectors provide a very useful tool.
A work of Hill, Lewandowski and Nurowski has given a good foundation for this,
reducing the field equations to a set of differential equations for two functions, one real, one complex, of three variables.
Under the assumption of the existence of one Killing vector,
the (infinite-dimensional) classical symmetries of those equations are determined and group-invariant solutions are considered.
This results in a single ODE of the third order which may easily be reduced to one of the second order.
A one-parameter class of power series solutions, $g(w)$, of this second-order equation is realized,
holomorphic in a neighborhood of the origin and behaving asymptotically as a simple quadratic function plus lower-order terms
for large values of $w$, which constitutes new solutions of the twisting type N problem.
The solution found by Leroy, and also by Nurowski, is shown to be a special case in this class.
Cartan's method for determining equivalence of CR manifolds is used to show that this class is indeed much more general.

In addition, for a special choice of a parameter, this ODE may be integrated once,
to provide a first-order Abel equation. It can also determine new solutions to the field equations
although no general solution has yet been found for it.
\end{abstract}

\pacs{04.20.Jb, 02.40.Tt, 02.30.Hq}

\submitto{\CQG}

\maketitle


\section{Introduction}
\label{introduction}

The search for general classes of solutions of the Einstein equations
that either have pure vacuum for their source or a non-zero cosmological constant, that are of Petrov type N,
and have principal null rays with non-zero twist has been continuing for a very long time now.
We will use the (relatively common) nomenclature ``Einstein spaces'' for solutions with this sort of a source,
i.e., either the pure vacuum or that vacuum with a non-zero cosmological constant, $\Lambda$, appended to it.
With this idea firmly in mind, there are only two solutions known so far for twisting type N Einstein spaces:
the one described by I. Hauser \cite{Hauser74,Hauser78}, which has $\Lambda=0$,
and the one by J. Leroy \cite{Leroy70}, with a non-zero value for $\Lambda$,
but which in the limit as $\Lambda\rightarrow 0$ degenerates to a flat solution, rather than one of type N.
Because of the apparent difficulty of the problem,
many different approaches have been used to attempt the finding of such a solution.
With a requirement of one or more Killing vectors, 
the problem can be reduced to the solution of a single, nonlinear ODE,
which has been produced in several forms by different authors \cite{Herlt86,McIntosh85,Chinea98,Finley94};
nonetheless this approach has produced no new solutions.
Looking at the problem as a reduction from complex-valued manifolds
via Pleba\'nski's hyperheavenly equation \cite{Plebanski76} has produced no new solutions \cite{Finley92}.
Therefore we were quite interested when we became aware of a different approach via a recent paper
by Pawe{\l} Nurowski \cite{Nurwoski08}, looking for exact solutions of this type with non-zero cosmological constant.

Many of Nurowski's research articles use the fact that one can productively study (4-dimensional) Lorentz geometries
which admit a shearfree, geodesic null congruence of curves by viewing the 3-parameter space
that picks out any particular curve in the congruence, as a (3-dimensional) CR structure \cite{Nurowski90}.
In \cite{Hill08}, he and his collaborators use the CR function in such a structure
to create a very appropriate choice of coordinates for a twisting type N Einstein space,
and reduce the Einstein equations to a set of nonlinear PDE's for a couple of functions of three variables.
Then in \cite{Nurwoski08}, he makes a clever ansatz depending only on a single variable and discovers a particular twisting solution;
unfortunately that solution turns out to be the same as the one mentioned above as first found by Leroy,
as he notes in a more recent paper \cite{Nurowski09}.
However, we were quite intrigued by the approach and have made some efforts to follow it through
with the hopes of obtaining more general solutions of the equations in Nurowski's article.

A spacetime of type N allows one, and only one, congruence of twisting, shearfree, null geodesics,
referred to as a principal null direction (PND);
this (3-parameter) family of null geodesics allows the option to choose a single coordinate $r$ along any such geodesic,
and to associate the other three degrees of freedom in the parameter space as a model for a CR manifold.
We first insist that our manifold admit a Killing vector in the real direction in this (3-dimensional) CR manifold,
so that the remaining unknown functions depend only on the complex coordinates there,
and then calculate the (infinite-dimensional) classical symmetries for the system.
This allows us to derive a quite simple nonlinear third-order ODE
which the invariant solutions of the classical symmetries must satisfy.
Because this equation does not contain the independent variable explicitly,
it can be immediately reduced to the following second-order ODE, for $g=g(w)$,
with two slightly different forms that differ by a constant:
\begin{equation} \label{geqn}
 g'' = - \frac{(g' + 2 w)^2}{2 g}  - \frac{2C}{g} - \frac{10}{3}, \qquad C = \hbox{$0$ or $1$.}
\end{equation}
We are then able to show that the Leroy-Nurowski solution is indeed a special solution for this equation.
At this point it is worthwhile to introduce a question as to how one knows
that the new solution of Nurowski does indeed describe locally\footnote{All our considerations are local,
both in the Lorentz-signature spacetime and in the associated complex spaces we need to use.}
the same manifold as the solution found by Leroy.
The method was originally created by E. Cartan \cite{Cartan32I,Cartan32II,Jacobowitz90} to prove equivalencies of CR structures,
without the need of actually determining an explicit coordinate transformation between the two sets of coordinates on two manifolds.
Instead, one determines the values of a set of invariant quantities for a CR structure, the same for all equivalent such structures.
Therefore it is necessary to calculate the invariants for Leroy's solution and
compare them with the ones already known to Nurowski for his solution, noting that they are the same constants.
We have therefore also calculated these invariants for our class of solutions, which we find to be quite different.

For the case when the constant $C$ in (\ref{geqn}) is zero,
an integral transformation may be performed to reduce that equation further, to a first-order ODE of Abel type, for $f=f(t)$:
\begin{equation*}
 f' = \frac{4}{t} \left(t+\frac{3}{2}\right) \left(t+\frac{1}{3}\right) f^3
 + \frac{5}{t} \left(t+\frac{2}{5}\right) f^2 + \frac{1}{2t} f,
\end{equation*}
which, quite unfortunately, we have not been able to identify
as any of the known solvable types of Abel equations \cite{Cheb-Terrab00,Cheb-Terrab03}.
Nonetheless, we believe that these two equations, at the time of publication,
are the simplest ODEs available that determine nontrivial twisting type N Einstein spaces.
Returning to the case when $C=1$, our current examinations suggest the possibility that
the solutions of this equation might define a new class of transcendental functions,
which constitutes the major result of this paper, establishing a new set of solutions to the type N problem with $\Lambda\neq 0$.
The remainder of the paper will describe the process involved in this and our reasons for stating that these are indeed new solutions.
In particular, we will present solutions to \eref{geqn} in the forms of power series and Puiseux series, both shown to be locally convergent.
Although prior to this time there were indeed only two known twisting type N Einstein spaces,
it is always worth remembering that the solution space for the problem is in fact quite large.
It has been shown by Sommers \cite{Sommers76} that the full set of solutions for type N is given by two complex functions of two real variables.
Surely the requirement of non-zero twist puts a very strong constraint on this, but it is expected
that there should be a large number of new analytic functions involved in the full solution of the twisting type N problem.

\section{CR structures and reduced Einstein equations}

As already mentioned those (4-dimensional) Lorentz geometries which admit a shearfree,
geodesic null congruence of curves have some history of people using CR structures to study them.
They were first introduced into mathematics by Poincar\'e and extensively studied by E. Cartan \cite{Cartan32I,Cartan32II}.
Good sources of background on the relation of these two geometric concepts may be found,
for instance in the thesis of Nurowski \cite{Nurowski93},
and also in the very detailed discussion of their use for Einstein spaces in his article with Hill and Lewandowski \cite{Hill08},
which generalized earlier work to include the possibility of a non-zero cosmological constant.

The approach to a Lorentz-signature manifold begins with the usual form for the metric \cite{Stephani03}
in terms of a basis of 4 null 1-forms, modified for our choice of signature, $(+,+,+,-)$:
\begin{equation}
 \mathbf g = 2 \left( \theta^1\theta^2 + \theta^3\theta^4 \right), \label{metric}
\end{equation}
where $\theta^2$ is the complex conjugate of $\theta^1$ while $\theta^4$ and $\theta^3$ are real,
and the product is the usual symmetric product of 1-forms to create the metric.
For Einstein spaces of type N, which have only one PND, we call its tangent vector $k$,
which is a well-defined vector field in the neighborhood of each point,
and take $\theta^3 = \mathbf{g}(k,\cdot)$ as its dual 1-form.
The requirement that the twist of $k$ should be non-zero
is equivalent to the restriction that $\theta^3\wedge \rmd\theta^3 \neq 0$.
We also choose a real coordinate $r$ as a parameter along this congruence,
and a real function $P\neq 0$ on the manifold so that
$k = P^{-1}\partial_r$ ($\partial_r \equiv \partial/\partial r$, $r\in \mathbb{R}$).
This congruence, still locally, is a 3-parameter family of null shearfree geodesics,
so that at each value of $r$, the manifold still needs three additional coordinates.
Following the work described in detail in \cite{Hill08}, one is able to describe those degrees of freedom
as coming from a 3-dimensional, strictly pseudoconvex CR manifold, $M$,
equipped with (at least) one non-constant CR function $\zeta$,
and classes of pairs of 1-forms $\lambda$ (real) and $\mu$ (complex) such that
\begin{equation}
 \eqalign{ \lambda \wedge \mu \wedge \bar{\mu} \neq 0, \qquad
 \mu = \rmd \zeta, \qquad \bar\mu  = \rmd \bar\zeta, \cr
 \rmd\lambda = \rmi\mu \wedge \bar\mu + (c\mu + \bar c \bar\mu )\wedge \lambda} \label{1-forms}
\end{equation}
with $c$ a complex-valued function on $M$.
A CR structure is a 3-dimensional real manifold $M$
equipped with an equivalence class of pairs of 1-forms, $(\lambda,\mu)$, as above.
Another pair $(\lambda',\mu')$ is considered equivalent to $(\lambda,\mu)$,
and therefore simply another representative of the same equivalence class,
iff there are functions $f\neq 0$ (real) and $h\neq 0$, $g$ (complex) on $M$ such that
\begin{equation} \label{CRETrfm}
 \lambda' = f\lambda, \qquad \mu' = h\mu + g\lambda, \qquad
 \bar\mu' = \bar h \bar\mu + \bar g \lambda.
\end{equation}
This CR manifold can be lifted into a spacetime,
so that the tetrad may be displayed in the following way \cite{Hill08}:
\begin{equation} \label{tetrad}
 \eqalign{ \theta^1 = P\,\mu, \qquad \theta^2 = P\,{\bar\mu},  \cr
 \theta^3 = P\,\lambda, \qquad \theta^4 = P \left(\rmd r + W\mu + \bar W \bar\mu + H\lambda \right). }
\end{equation}
Of course this is exactly the form that a spacetime with such a distinguished PND is supposed to have \cite{Stephani03}.

Referring back to \eref{tetrad}, we can see that the twist is just proportional to $\lambda \wedge \rmd \lambda$.
Using \eref{1-forms} to determine this, we see that
these forms have been chosen to put the non-zero character of the twist very clearly in evidence as non-zero.
However, still looking at those equations,
the function $c$ that appears there is quite an important new function on the manifold.
Using the closure of the equations in \eref{1-forms}, one determines an important reality condition on the derivatives of $c$:
\begin{equation*}
 \partial\bar c =  \bar\partial c,
\end{equation*}
once a dual basis of vector fields is introduced,
which, however, is not a commutative basis:
\begin{eqnarray*}
 \left(\partial_0, \partial, \bar\partial \right) \hbox{~~dual to~~} \left(\lambda, \mu, \bar\mu \right), \\
 \left[\partial, \bar\partial\, \right] = -\rmi \partial_0, \qquad
 \left[\partial_0, \partial \right] = c \partial_0, \qquad
 \left[\partial_0, \bar\partial\, \right] = \bar c \partial_0.
\end{eqnarray*}
At this point one has sufficient information to write out the explicit forms of the Einstein
equations, which, in generality, say that $R_{12} = \Lambda = R_{34}$ with all the other components of the
Ricci tensor vanishing.  We quote from \cite{Hill08} and \cite{Nurwoski08} which show that the
results are the following:
\begin{equation*}
 P = \frac{p}{\cos(\case{r}{2})}, \qquad W = \rmi\,a\,(1+\rme^{-\rmi r}), \qquad
 H = q\,\rme^{\rmi r} + \bar q \,\rme^{-\rmi r} + h,
\end{equation*}
where the functions $a$, $q$ (complex) and $h$, $p$ (real), all independent of $r$, satisfy
\begin{eqnarray*}
 a = c + 2 \partial \log p, \\
 q = \frac{2}{3} \Lambda p^2 + \frac{2 \partial p\, \bar\partial p
 - p\, \left(\partial \bar\partial p + \bar\partial \partial p \right)}{2 p^2} - \frac{\rmi}{2}\,
 \partial_0 \log p - \bar\partial c, \\
 h = 2 \Lambda p^2 + \frac{2 \partial p\, \bar\partial p
 - p\, \left(\partial \bar\partial p + \bar\partial \partial p \right)}{p^2} - 2 \bar\partial c.
\end{eqnarray*}
Given all the above, the functions $a,c,h,p$ and $q$ define a twisting type N Einstein space,
of the form given in \eref{metric} and \eref{tetrad},
iff the unknown functions $c$ and $p$ satisfy the following system of PDEs on $M$
\begin{eqnarray}
 \partial \bar c = \bar\partial c \label{TN1} \\
 \left[ \partial
 \bar\partial + \bar\partial \partial + \bar c
 \partial + c \bar\partial + \case{1}{2} c \bar c + \case{3}{4}
 \left(\partial \bar c + \bar\partial c \right) \right] p =
 \case{2}{3} \Lambda p^3, \label{NurowskiEqn} \\
 \Psi_3 = 0, \qquad R_{33} = 0. \label{TN3}
\end{eqnarray}
as well as one inequality
\begin{equation*}
 \Psi_4 \neq 0,
\end{equation*}
in order that the spacetime should not be conformally flat.
In terms of those variables already defined, the Ricci tensor component $R_{33}$ and the Weyl scalars $\Psi_3$ and $\Psi_4$
take on quite nasty-looking expressions
\begin{eqnarray*}
 R_{33} =& \Bigg[ \frac{8}{p^4}
 \left(\partial + 2 c \right) \left(p^2 \partial \bar I \right)
 -8 \Lambda \Big( \case{4}{3} \Lambda p^2
 + 6 \left(\bar c \partial + c \bar\partial \right) \log p \\
 & + 12 \partial \log p \; \bar\partial \log p + 3 c \bar c
 - \bar\partial c - 2 \rmi \partial_0 \log p \Big) \bigg] \cos^4 \left(\frac{r}{2} \right),
\end{eqnarray*}
\begin{eqnarray*}
 \Psi_3 = \left[ \frac{2\rmi}{p^2}\ \partial\bar I
 - 4 \rmi \Lambda \left(2 \bar\partial \log p + \bar c
 \right) \right] \mathrm{e}^{\rmi r/2} \cos^3 \left(\frac{r}{2} \right),
\end{eqnarray*}
\begin{eqnarray*}
 \fl \Psi_4 = \left\{ \frac{2\rmi}{p^2}\ \partial_0 \bar I
 + \frac{4}{3} \Lambda \left[ \left(\bar\partial + \bar c \right)
 \left(2\bar\partial \log p + \bar c \right) + 2 \left(2\bar\partial
 \log p + \bar c \right)^2 \right] \right\} \mathrm{e}^{-\rmi r/2} \cos^3
 \left(\frac{r}{2} \right),
\end{eqnarray*}
where the function $I$ is defined by
\begin{equation*}
 I = \partial \left(\partial \log p + c \right) + \left(\partial \log p + c \right)^2.
\end{equation*}
Following the procedure of \cite{Hill08} (with the use of Maple)
which calculated the curvature tensor using Cartan's structure equations,
we present the calculated $\Psi_4$ above (\emph{simplified} with the use of $\Psi_3=0$) for $\Lambda\neq 0$,
which we believe has not been previously published.
Despite the frightening appearance of $R_{33}$,
the equations \eref{NurowskiEqn} and $\Psi_3=0$ together do imply the requirement $R_{33}=0$.
This tells us that within the established formalism
the twisting type N solutions to the Einstein equations automatically satisfy the condition for an Einstein space,
i.e., vacuum with or without a cosmological constant.
For $\Lambda=0$, the statement is obviously true (see also \cite{Stephani03} p.451) and was used in \cite{Hill08}
to prove the CR embeddability of twisting type N vacuums, without cosmological constant.
For $\Lambda \neq 0$, one uses \eref{NurowskiEqn} to substitute the term $\frac{4}{3} \Lambda p^2$ in $R_{33}$
and notices that the resulting expression is a linear combination of $\partial \Psi_3$ and $\Psi_3$.
The equation $R_{33}=0$ is therefore superfluous for the type N problem,
which facilitates our calculation greatly.

\section{Killing vector in the $u$-direction}

It is useful to understand the meaning of the operator $\partial$
by introducing a real coordinate system $(x, y ,u)$ on $M$ such that we have
\begin{equation}
 \!\!\! \begin{array}{ll}
 \zeta = x + \rmi y, \qquad & \partial_\zeta = \case{1}{2} \left(\partial_x - \rmi \partial_y \right), \\
 \partial = \partial_\zeta - L \partial_u,
 \qquad & \partial_0 = \rmi(\bar\partial L - \partial\bar{L}) \partial_u,
 \end{array} \qquad
 \lambda = \frac{\rmd u + L \rmd\zeta + \bar{L} \rmd\bar\zeta}{\rmi(\bar\partial L - \partial\bar{L})}, \label{lambda}
\end{equation}
with $L=L(\zeta, \bar\zeta, u)$ a complex-valued function \cite{Hanges88}.
In addition, the function $L$ relates to the function $c$ in the following way
\begin{equation}
 c = -\partial \ln (\bar\partial L - \partial\bar{L}) - \partial_u L. \label{candL}
\end{equation}

One major difficulty of fully solving the system (\ref{TN1}-\ref{TN3}) is that
unlike ordinary coordinate differentiations, the selected (dual) basis for the tangent space is not commutative,
and, even worse, the operator $\partial$ itself involves the unknown function $c$.
When this dependence on the real coordinates is written out explicitly, the original PDEs become formidably lengthy.
Instead of facing this entire conundrum, we have decided to circumvent it, at least in this paper,
by looking at the special case where the unknowns $p$ and $c$ have no $u$-dependence, i.e., $\partial_0 p = 0 = \partial_0 c$.
Geometrically speaking, we insist that the spacetime admits a Killing vector in the $u$-direction.
Such an assumption simplifies the problem greatly in that
one can treat the operator $\partial$ the same as $\partial_\zeta$, when acting on either $p$ or $c$.
This is a generalization of the assumption made by Nurowski \cite{Nurwoski08},
where it was simplified to just dependence on $y$, i.e, two Killing vectors assumed.

\textbf{Theorem 1} (\emph{CR embeddability} \cite{Nurowski88})
A CR structure \eref{1-forms} with $c=c(\zeta, \bar\zeta)$ is CR embeddable.
From this particular form of $c$, a $u$-independent form of the function $L$ can be constructed as
\begin{equation}
 L(\zeta,\bar\zeta)
 = -\case{\rmi}{2} \int \alpha(\zeta,\bar\zeta)\, \rmd \bar{\zeta} \label{Lfun}
\end{equation}
with a real-valued function $\alpha \neq 0$ satisfying
\begin{equation}
 \partial_\zeta \alpha = -c\, \alpha, \qquad \partial_{\bar\zeta} \alpha = -\bar{c}\, \alpha. \label{alpha}
\end{equation}
Associated to this $L$, the tangential CR equation $\bar\partial \eta=0$ yields a second CR function:
\begin{equation}
 \eta = u + \case{\rmi}{2} \int \!\!\!\! \int \alpha(\zeta,\bar\zeta)\, \rmd
 \zeta \mathrm{d} \bar{\zeta}. \label{2CRfun}
\end{equation}

\emph{Proof:} Because of the restraint $\partial_\zeta \bar c = \partial_{\bar\zeta} c$,
the system \eref{alpha} is compatible and has a real solution $\alpha\neq 0$.
Henceforward one can directly check that \eref{Lfun} satisfies \eref{candL} and that
\eref{2CRfun} satisfies the equation $\bar\partial \eta=(\partial_{\bar\zeta} - \bar{L} \partial_u) \eta =0$.
Clearly, the CR functions $\eta$ and $\zeta$ are functionally independent, i.e., $\rmd \zeta \wedge \rmd \eta \neq 0$.
Therefore we acquire a second CR function.

For a given function $c=c(\zeta,\bar\zeta)$, the equation \eref{candL}, viewed as a PDE for $L$,
may give rise to multiple choices of the function $L$, hence various $\lambda$'s.
However, such an ambiguity only constitutes different representatives of the same CR structure.
To see this, one may look into the six Cartan invariants (details in Section \ref{CRsymmetry}) and notice that
they are all uniquely determined by the function $c=c(\zeta,\bar\zeta)$ (see, e.g., \eref{alphaI} and Appendix B),
given that the function $r$ defined in \eref{alphaI} does not vanish.
For CR structures with $r=0$, they are all locally equivalent to a 3-dimensional hyperquadric inside
$\mathbb{C}^2$ \cite{Cartan32I,Jacobowitz90}.
An alternative proof would be to show that there always exists a coordinate transformation
$u \rightarrow \tilde{u}(\zeta,\bar\zeta,u)$ ($\partial\tilde{u}/\partial u \neq 0$)
that takes a function $L=L(\zeta,\bar\zeta,u)$ satisfying \eref{candL} to the $u$-independent form \eref{Lfun}.
This can be confirmed by checking the compatibility of PDEs regarding such an existence.
In conclusion, the CR structure on $M$ is \emph{uniquely} determined once a function $c=c(\zeta,\bar\zeta)$ is given.

The converse of the last statement above is, however, not true.
In fact various choices of the function $c$ may correspond to the same CR structure.
We will see examples of this in later sections.
Related to this matter, our assumption of the function $c$ being $u$-independent is thus not a CR invariant property.
A function $c=c(\zeta,\bar\zeta)$ may acquire $u$-dependence through the transformation \eref{CRETrfm}
that takes one representative $(\lambda,\mu)$ of the CR structure to another.

Now we apply the assumption and the following notations
\begin{equation*}
 \partial f \rightarrow \partial_\zeta f = f_1, \ \bar\partial f \rightarrow \partial_{\bar\zeta}f = f_2, \
 f=f_0, \ (f=p,\ c\ \textrm{and}\ \bar{c}\ \textrm{only})
\end{equation*}
and then rewrite the system (\ref{TN1}-\ref{TN3}) as
\begin{eqnarray}
 \bar{c}_1 = c_2, \label{Cc}\\
 2 p_{12} + \bar{c}_0 p_1 + c_0 p_2 + \case{1}{2} c_0 \bar{c}_0 p_0 + \case{3}{4}(\bar{c}_1+c_2) p_0
 = \case{2}{3}\Lambda p_0^3, \label{NurowskiEqn1} \\
 \fl p_0 p_{122} - p_1 p_{22} + 2 \bar{c}_0 p_0 p_{12} - 2\bar{c}_0 p_1 p_2 + 2 \bar{c}_1 p_0 p_2
 + (\bar{c}_{12} + 2\bar{c}_0 \bar{c}_1) p_0^2 = 2\Lambda (2p_2+\bar{c}_0 p_0) p_0^3, \label{Psi3}
\end{eqnarray}
where the last equation arises from $\Psi_3=0$.
These are the PDEs we aim to solve. Moreover, the Weyl scalar $\Psi_4$ reads
\begin{equation}
 \Psi_4 = \frac{4}{3} \Lambda \left[
 2p_0 p_{22} + 6p_2^2 + 10 \bar{c}_0 p_0 p_2 + (\bar{c}_2 + 3\bar{c}_0^2) p_0^2
 \right] \frac{\mathrm{e}^{-\rmi r/2}}{p_0^2} \cos^3\left(\frac{r}{2} \right). \label{Psi4}
\end{equation}

\section{Infinite-dimensional classical symmetries}

We follow the standard procedure (e.g., \cite{Krasilshchik99})
to calculate the classical symmetries of the system (\ref{Cc}-\ref{Psi3}).
Since \eref{Psi3} is generally complex, we have to include its complex conjugate as well in the calculation.
Moreover, we treat \eref{Cc} as a constraint
and encode it and its differential consequences directly into the choice of intrinsic coordinates
so that this equation no longer needs further attention.
This gives us three PDEs for three dependent variables $p$, $c$ and $\bar c$
which depend on two independent variables $\zeta$ and $\bar{\zeta}$.
The intrinsic coordinates within the first four jets that are relevant to the calculation are chosen as follows
\begin{eqnarray*}
 p_0,c_0,\bar{c}_0, \\
 p_1,p_2,c_1,\bar{c}_1,\bar{c}_2, \\
 p_{11},p_{22},c_{11},\bar{c}_{22}, \\
 p_{111},p_{222},c_{111},\bar{c}_{222}.
\end{eqnarray*}
The rest of the jet variables, such as $p_{12}$, $p_{122}$, $\bar{c}_{12}(=c_{22})$ etc., can be expressed
in terms of the intrinsic coordinates through the PDEs and their differential consequences.

With a considerable amount of manual work on the algebraic computer program Maple,
we have managed to find the classical symmetries with the generating section given by
\begin{eqnarray*}
 \Psi &=& -\case{1}{2} (\partial_\zeta A + \partial_{\bar{\zeta}} \bar{A}) p_0
 - A p_1 - \bar{A} p_2, \\
 \Theta &=& \partial^2_\zeta A - (\partial_\zeta A) c_0 - A c_1 -
 \bar{A} \bar{c}_1, \\
 \bar{\Theta} &=& \partial^2_{\bar{\zeta}} \bar{A} - (\partial_{\bar{\zeta}} \bar{A}) \bar{c}_0
 - A \bar{c}_1 - \bar{A} \bar{c}_2,
\end{eqnarray*}
where $A=A(\zeta)$ is an arbitrary function of $\zeta$ that is sufficiently differentiable.
The Lie bracket of two symmetries with, respectively, $A_1(\zeta)$ and $A_2(\zeta)$
yields a third symmetry with a new $A_3(\zeta)$ given by
\begin{equation*}
 A_3 = [A_1,A_2] := A_1 \partial_\zeta A_2 - A_2 \partial_\zeta A_1.
\end{equation*}
Therefore, we indeed obtain an infinite-dimensional set of classical symmetries for the system (\ref{Cc}-\ref{Psi3}).
In particular, they reduce to translational symmetries for nonzero constant $A$,
and scaling symmetries for $A \propto\zeta$.

\section{Group-invariant solutions and reductions to ODEs}

Setting the generating section $(\Psi,\Theta,\bar{\Theta})$ to zero and solving for $p$, $c$ and
$\bar c$, we are able to obtain a remarkable ansatz for the Einstein equations:
\begin{equation}
 \fl p(\zeta, \bar{\zeta}) = \frac{F_1(z)}{\sqrt{A \bar{A}}}, \quad
 c(\zeta, \bar{\zeta}) = \frac{\partial_\zeta A + \rmi F_2(z) + C_1}{A}, \quad
 \bar{c}(\zeta, \bar{\zeta}) = \frac{\partial_{\bar{\zeta}} \bar{A}
 - \rmi F_2(z) + C_1}{\bar{A}} \label{pccb}
\end{equation}
with a new real argument
\begin{equation}
 z = -\rmi \left( \int \frac{1}{A} \mathrm{d} \zeta -
 \int \frac{1}{\bar A} \mathrm{d} \bar{\zeta} \right)
 = \mathrm{Im} \int \frac{2}{A} \mathrm{d} \zeta, \label{zdef}
\end{equation}
where the constant $C_1$ and the undetermined functions $F_{1,2}(z)$ are all real-valued.
One may easily verify these expressions by direct calculation.

Substituting the ansatz into \eref{NurowskiEqn1} and \eref{Psi3} and noticing
that all dependence on $A, \bar{A} \neq 0$, except those in the argument $z$, can be factored out,
we have a neat reduction from the PDEs to a system of two ODEs for $F_1$ and $F_2$ only:
\begin{eqnarray*}
 0 = -F_1^{\prime \prime} + F_2 F_1' + \case{1}{3}\Lambda F_1^3 -
 \case{1}{4} (F_2^2 - 3 F_2' + C_1^2) F_1, \\
 0 = - F_1^{\prime \prime} F_1 + (F_1')^2 + \Lambda F_1^4 + F_2'
 F_1^2.
\end{eqnarray*}
The satisfaction of the second ODE above is given
by the introduction of a single new, real-valued function $J=J(z)$ such that
\begin{equation}
 F_1 = \pm \sqrt{J'}, \qquad F_2 = \frac{J^{\prime\prime}}{2 J'} - \Lambda J. \label{F12J}
\end{equation}
Then the first ODE simply becomes
\begin{equation}
 J''' =
 \frac{(J^{\prime\prime})^2}{2J'} - 2 \Lambda J J^{\prime\prime} -
 \frac{10}{3} \Lambda (J')^2 - 2 (\Lambda^2 J^2 + C_1^2) J'. \label{JEQ}
\end{equation}
Since this ODE does not have the argument $z$ appearing explicitly,
we can lower the order of the ODE through the standard transformation
\begin{equation*}
 J' = P(J) \Longrightarrow J''=P P' \Longrightarrow J'''= P(PP')'
\end{equation*}
and obtain an even simpler equation of the second-order
\begin{equation} \label{PEQ}
  P'' = - \frac{(P' + 2 \Lambda J)^2}{2P} - \frac{2 C_1^2}{P} - \frac{10}{3} \Lambda.
\end{equation}
A solution $P=P(J)$ to \eref{PEQ} can give rise to a solution
$J=J(z)$ to \eref{JEQ} at least locally by inverting
\begin{equation} \label{PJz}
 z + C_0 = \int \frac{1}{P(J)} \mathrm{d} J
\end{equation}
with $C_0$ constant. This solution will be physical
if it also makes $F_{1,2}(z)$ real-valued via \eref{F12J}, which requires that locally
\begin{equation}
 P(J) > 0, \ J' > 0 \ \textrm{and}\ J\ \textrm{real-valued}.
\end{equation}
Therefore we are only interested in solutions for $J(z)$ that are monotonically increasing,
or equivalently, positive $P(J)$.

We can also consider the special case of \eref{PEQ} with $C_1=0$ and $\Lambda\neq 0$, i.e.,
\begin{equation} \label{PEQA}
 P'' = - \frac{(P' + 2 \Lambda J)^2}{2P} - \frac{10}{3} \Lambda.
\end{equation}
By introducing the following integral transformation
\begin{equation*}
 J = \frac{1}{\Lambda} \exp\left( \int f(t)\ \mathrm{d}t \right),
 \qquad P(J) = \frac{t}{\Lambda} \exp \left( 2 \int f(t)\ \mathrm{d}t \right),
\end{equation*}
of which the inverse has the form
\begin{equation*}
 t = \frac{P}{\Lambda J^2}, \qquad f(t) = \frac{\Lambda J^2}{J P' - 2 P},
\end{equation*}
we can further reduce \eref{PEQA} to an Abel ODE of the first kind \cite{Murphy60},
as already noted in the Introduction:
\begin{equation} \label{Abel}
 f' = \frac{4}{t} \left(t+\frac{3}{2}\right)\left(t+\frac{1}{3}\right) f^3
 + \frac{5}{t} \left(t+\frac{2}{5}\right) f^2 + \frac{1}{2t}\ f.
\end{equation}
Once the general solution $f=f(t, C_2)$ is acquired with a constant $C_2$,
we can find the general solution $P(J)$ of \eref{PEQA} by solving the following ODE
\begin{equation*}
 f\!\left(\frac{P}{\Lambda J^2}, C_2\right) = \frac{\Lambda J^2}{JP'-2P},
\end{equation*}
of which the solution is given by
\begin{equation} \label{PTNA}
 P(J) = Z(J) J^2, \hbox{~~with~~} 0 = -\ln J + \int^{Z/\Lambda} \!\!\!\! f(t, C_2)\ \mathrm{d}t + C_3.
\end{equation}

Simple as both \eref{PEQ} and \eref{Abel} may appear,
so far we have had no luck finding their explicit general solutions.
For more comments on \eref{Abel} and Abel ODEs in general, see Appendix A.

\section{CR equivalency as classical symmetry}
\label{CRsymmetry}

To identify new twisting type N Einstein spaces obtained from \eref{JEQ},
we refer to the following theorem as a natural way to classify metrics equipped with CR structures.

\textbf{Theorem 2} (\cite{Hill08}, see \emph{Theorem 1.2} and references therein) Let $(\mathcal{M},g)$ be a 4-dimensional manifold
equipped with a Lorentzian metric and foliated by a 3-parameter family of shearfree null geodesics.
Then $\mathcal{M}$ is locally a Cartesian product $\mathcal{M} = M \times \mathbb{R}$.
The CR structure $(M, (\lambda, \mu))$ on $M$ is uniquely determined by $(\mathcal{M}, g)$
and the shearfree null congruence on $\mathcal{M}$.

By definition, a type N spacetime at each point has a unique PND.
In the case of vacuums (with or without $\Lambda$), a PND must be geodesic and shearfree \cite{Goldberg62}.
Thus for every twisting type N Einstein space, the shearfree null congruence is unique.
Hence to confirm a new twisting type N vacuum metric,
it is sufficient to show that its CR structure is distinct from the one of known metrics.
This can be routinely done by computing the six Cartan invariants \cite{Cartan32I,Cartan32II,Jacobowitz90},
which are denoted respectively by
\begin{eqnarray}
 \alpha_I, \theta_I, \eta_I\ \textrm{(complex)}, \nonumber \\
 \beta_I, \gamma_I, \zeta_I\ \textrm{(real)}. \nonumber
\end{eqnarray}
Cartan showed that two local CR structures are equivalent
iff their six CR invariants (defined when $r\neq 0$ in \eref{alphaI}) are identical,
except possibly for a sign difference in both $\alpha_I$ and $\eta_I$ \cite{Cartan32I}.
With the assumption of $u$-independence,
we can write down, for instance, the simplest invariant computed from the 1-forms defined in \eref{1-forms}:
\begin{eqnarray}
 \eqalign{ \alpha_I(\zeta,\bar\zeta) = -\frac{5 \bar{r} \partial_\zeta r + r \partial_\zeta \bar{r} + 8 c r\bar{r}}
{8 \sqrt{\bar{r}} \cdot \sqrt[8]{(r\bar{r})^{7}}}, \cr
 r = \case{1}{6}\left(\partial_{\bar\zeta} \bar{l} + 2 \bar{c} \bar{l}\right), \qquad
 l = -\partial_\zeta \partial_{\bar\zeta} c - c \partial_{\bar\zeta} c.} \label{alphaI}
\end{eqnarray}
Here the function $r\neq 0$, following the notation of Cartan,
is not to be confused with the coordinate $r$ along the null congruence.
For our calculated $\beta_I, \gamma_I$ and $\theta_I$, see Appendix B.
Note that this $\alpha_I$ only relies on $c(\zeta,\bar\zeta)$, $\bar{c}$ and their derivatives,
which is also the case for all the other Cartan invariants.
We first point out a remarkable feature of these invariants computed from the ansatz (\ref{pccb},\ref{zdef}).

\textbf{Proposition} Given the ansatz (\ref{pccb},\ref{zdef}),
all the following quantities are independent of $A(\zeta)$ and $\bar{A}(\bar\zeta)$ except those in the argument $z$:
\begin{equation*}
 \alpha_I^2, \eta_I^2, \alpha_I \bar\eta_I, \beta_I, \gamma_I, \theta_I, \zeta_I.
\end{equation*}
In another word, they are all functions of $z$ only, e.g., $\beta_I(\zeta,\bar\zeta)=\beta_I(z)$.

\emph{Proof}: Except for a lengthy but straightforward symbolic computation with Maple,
we are, at the moment, still not aware of any other more insightful way of proving this result.
Here we only emphasize that the law $\sqrt{v}\sqrt{w}=\sqrt{vw}$ is in general \emph{not} true in the complex domain;
failing to notice this may cause an erroneous conclusion.

\textbf{Remark} For a fixed $z$,
the presence of the functions $A$ and $\bar{A}$ in $\alpha_I$ and $\eta_I$ themselves only affects their signs.
More specifically, the only dependence on $A$ and $\bar{A}$ takes the following forms:
\begin{eqnarray*}
 \alpha_I \propto \frac{1}{A(\zeta)} \sqrt{\frac{A^2(\zeta)}{F(z)}}, \qquad
 \eta_I \propto \frac{1}{\bar{A}(\bar\zeta)} \sqrt{\frac{\bar{A}^2(\bar\zeta)}{\bar F(z)}}, \\
 F(z) = -F_2^{\prime\prime\prime} + (F'_2)^2 + 3F_2 F_2^{\prime \prime} - 2F_2^2 F'_2
 + 2 C_1^2 F'_2 + \rmi C_1 (3F_2^{\prime \prime} - 4F_2 F'_2).
\end{eqnarray*}
Hence the product $\alpha_I \bar\eta_I$ is a function of $z$ only.
According to Cartan \cite{Cartan32I},
this sign situation is accounted for by a local CR diffeomorphism and therefore does not generate a new CR structure.
Hence we have proved the following theorem.

\textbf{Theorem 3} Locally, the CR structure \eref{1-forms} (as an equivalence class) determined by the function $c$ given in (\ref{pccb},\ref{zdef})
is independent of the choice of the function $A(\zeta)\neq 0$, once the form of $F_2(z)$ is fixed.

Altogether, the freedom of choosing various $A(\zeta)\neq 0$ does not affect the CR structure of a type N metric which we are considering.
Hence, for the simplicity of representing new metrics distinguished by CR structure,
we can just set $A(\zeta)=\bar{A}(\bar{\zeta})=2$ (see the Conclusions).
In hindsight, the classical symmetries we have obtained are nothing more than a particular manifestation of the underlying CR equivalency.
We believe this connection between the two may as well suggest a more general concern
if one aims to find, through the (classical or higher) symmetries,
additional exact solutions to the Einstein equations formulated with CR structures.

We will see later examples of solutions that have \emph{constant} CR invariants,
and remarkably, one of them is the solution of Leroy-Nurowski.
Nonetheless, this feature is generally not true for other solutions.

\section{Conformally flat solutions}
Before we try to solve \eref{JEQ} for type N solutions,
it is important to find out in advance those conformally flat solutions satisfying $\Psi_4=0$
which are automatically contained in the general solution of \eref{JEQ}.
We insert the ansatz (\ref{pccb},\ref{zdef}) into the expression for $\Psi_4$ given by \eref{Psi4},
and re-normalize $\Psi_4$ to pull out just a simple complex-valued function of $z$:
\begin{eqnarray}
 \fl K(z) &:= -\frac{3\bar{A}^2 F_1^2 \mathrm{e}^{\rmi r/2}}{4\Lambda
 \cos^3 \left(\frac{r}{2} \right)} \Psi_4 \nonumber \\
 \fl &= 2 F_1 F_1''
 + 6 (F_1')^2 - 10 (F_2 + \rmi C_1) F_1 F_1' + (-F_2' + 3 F_2^2 + 6 \rmi C_1 F_2 -
 3 C_1^2) F_1^2.
\end{eqnarray}
We now apply \eref{F12J} and use \eref{JEQ} to substitute for $J'''$, which gives us
\begin{equation} \label{KEQ}
 K = \left[\Lambda J J'' - \case{2}{3}\Lambda (J')^2 + 2(\Lambda^2 J^2 - 2 C_1^2) J'\right]
 + \rmi \left[ -2 C_1 \left(J'' + 3 \Lambda J J' \right) \right],
\end{equation}
or in terms of $P(J)$,
\begin{equation*}
 K = P \left[\Lambda J P' -\case{2}{3} \Lambda P + 2 \Lambda^2 J^2 - 4 C_1^2 \right]
 + \rmi \left[ -2 C_1 P \left(P' + 3\Lambda J \right) \right],
\end{equation*}
where we have put the real and imaginary parts in separate brackets.
Replacing $J''$ with the help of \eref{KEQ}, we can rewrite the equation \eref{JEQ} as
\begin{equation*}
 0 = \case{1}{3} \Lambda K (J')^2 - (2\Lambda K J + K')(\Lambda J - 2\rmi C_1) J' + \case{1}{2} K^2
\end{equation*}
which clearly has $K=0$, i.e, all conformally flat solutions, as some of its solutions.

If $P(J)$ is not restricted to the real domain,
then solving the first-order ODE $K=0$ for $P(J)$ leads to the following general solution
\begin{equation} \label{PFlat}
 P(J) = - \frac{3}{2} \Lambda \left(J^2 + \frac{4C_1^2}{\Lambda^2} \right)
 + C_2 \left(J \pm \frac{2\rmi C_1}{\Lambda}\right)^{2/3}.
\end{equation}
with a complex constant $C_2$.

If, instead, we restrict $P(J)$ to be real,
a simultaneous vanishing of the real and imaginary parts of $K$ respectively
yields the following set of two equations, provided $P\neq 0$,
\begin{equation*}
 \eqalign{0 = C_1 (P' + 3 \Lambda J), \cr
 P' = \frac{2P}{3J} - 2 \Lambda J + \frac{4 C_1^2}{\Lambda J},}
\end{equation*}
both of which are consistent with \eref{PEQ}. There are now two cases for solutions.

The case $C_1 \neq 0$ requires that both ODEs be satisfied, so that we have a unique solution
\begin{equation} \label{PFlat1}
 P(J) = -\frac{3}{2}\Lambda J^2 - \frac{6 C_1^2}{\Lambda}
\end{equation}
which, by solving $J'=P(J)$, gives rise to
\begin{equation} \label{JFlatS}
 J = \frac{2 C_1}{\Lambda \tan (3 C_1 (z + C_0))}.
\end{equation}
In the limit $C_1 \rightarrow 0$, the above solution becomes even simpler\footnote{
Both \eref{JFlatS} and \eref{JFlatSS} would be of particular importance for the perturbation theory
on type N solutions near flat ones.}:
\begin{equation} \label{JFlatSS}
 J = \frac{2}{3 \Lambda (z + C_0)}.
\end{equation}
From \eref{JFlatS}, we have
\begin{equation}
 F_1 = \pm \frac{\sqrt{6} C_1}{s \sin(3C_1 (z + C_0))}, \qquad
 F_2 = -\frac{5 C_1}{\tan(3 C_1 (z + C_0))}
\end{equation}
with negative-valued $\Lambda = -s^2$.
Note that it is only at this stage that the reality condition on $F_{1,2}$, i.e., $J'>0$,
requires $\Lambda < 0$, i.e, a negative cosmological constant.
An important remark that can be made is that the extended form of the Leroy-Nurowski solution (see the next section)
resembles this solution greatly with simply differences in the coefficients.

For the other case when $C_1 = 0$, we have
\begin{equation} \label{PFlat2}
 P(J) = - \case{3}{2} \Lambda J^2 + C_2 J^{2/3}.
\end{equation}
with a real constant $C_2$.
From \eref{PJz}, the solution $J(z)$ is determined by
\begin{equation} \label{JFlatG}
 \int \frac{1}{- \case{3}{2} \Lambda J^2 + C_2 J^{2/3}}
 \mathrm{d} J = z + C_0.
\end{equation}
Since $J'>0$, we cannot have both $\Lambda>0$ and $C_2\leq 0$.
Hence, we can discuss three other sign possibilities, the details of which are put in Appendix C.

We note that the special solution \eref{JFlatSS} corresponding to $C_1=C_2=0$ serves as the single ``point''
where these two families of conformally flat solutions are joined up.

Modulo possible sign differences in $\alpha_I$ and $\eta_I$ caused by square roots as already discussed,
the Cartan invariants for both \eref{JFlatS} and \eref{JFlatSS}, as calculated via \eref{alphaI}
and the equations for the other invariants, as presented in Appendix B, are given by
\begin{eqnarray}
 \eqalign{ \alpha_I = -\frac{4\rmi}{\varepsilon} \sqrt[4]{\frac{2}{5}}, \qquad
 \beta_I = \frac{41}{2\sqrt{10}}, \qquad
 \gamma_I = \frac{29}{2 \sqrt{10}}, \cr
 \theta_I = 3\rmi \sqrt{\frac{2}{5}}, \qquad
 \eta_I = -\frac{\rmi}{\varepsilon} \cdot \frac{2^{19/4}}{5^{3/4}}, \qquad
 \zeta_I = -\frac{327}{40}, \qquad \varepsilon=\pm 1.} \label{CartanInvarPFlat}
\end{eqnarray}
Remarkably, they are all constant and do not depend on $C_1$.
Nonetheless, this is not the case for the other conformally flat solutions obtained from \eref{JFlatG} with $C_2\neq 0$
of which the Cartan invariants are generally functions of $z$ and $C_2$.
For instance, simplified by \eref{PFlat2} and $J'=P(J)$, the first Cartan invariant satisfies
\begin{equation} \label{CartanInvarPFlat2}
 \alpha_I^2(z,C_2) = -16 \sqrt{\frac{2}{5}} \left(\frac{3\Lambda J^{4/3} + 2C_2}{3\Lambda J^{4/3} - 2C_2}\right)^2,
\end{equation}
where $J=J(z)$ belongs to one of the three cases described in Appendix C.

Two conformally flat Einstein spaces may have non-equivalent CR structures.
This does not conflict with the previous theorem because in a conformally flat spacetime,
one is free to make different choices from among the multiple shearfree null congruences
and therefore may have non-equivalent CR structures attached to them.

\section{An extended form of the Leroy-Nurowski solution}

Now we can reveal a fuller extent of the exact twisting type N solution first discovered by Leroy,
and re-derived by Nurowski within the framework of CR geometry,
upon the latter of which our current work is mainly based.
We hope that our derivation of this solution will make the process behind the previous discoveries appear clearer.

Given Nurowski's form of the solution (see \cite{Nurwoski08} or \eref{Nurowskisoln})
and recasting it into the form of the ansatz (\ref{pccb},\ref{zdef}) and \eref{F12J},
we find the following special solution to \eref{PEQ}
\begin{equation} \label{PTN}
 P(J) = -\frac{1}{3}\Lambda J^2 - \frac{3 C_1^2}{4\Lambda}
\end{equation}
which gives rise to a solution to \eref{JEQ}:
\begin{equation} \label{JTN}
 J = \frac{3 C_1}{2 \Lambda \tan (\frac{1}{2} C_1 (z + C_0))}.
\end{equation}
In the limit $C_1 \rightarrow 0$, the above expression becomes even simpler:
\begin{equation} \label{JTNS}
 J = \frac{3}{\Lambda (z + C_0)},
\end{equation}
which is quite similar to that of \eref{JFlatSS}.
Back to the case with $C_1\neq 0$, using \eref{F12J}, we have
\begin{equation}
 F_1 = \pm \frac{\sqrt{3} C_1}{2 s \sin(\frac{1}{2} C_1 (z + C_0))},
 \qquad F_2 = -\frac{2 C_1}{\tan(\frac{1}{2} C_1 (z + C_0))}
\end{equation}
with a negative $\Lambda = -s^2$.
Note that it is only at this stage that the reality condition on $F_1$ requires $\Lambda < 0$.
In the end, our extended version of the Leroy-Nurowski solution takes the form
\begin{eqnarray}
 p(\zeta, \bar{\zeta}) = \pm \frac{\rmi \sqrt{3} C_1}{2 s \sinh \left(\frac{\rmi}{2} C_1
 (z + C_0) \right) \sqrt{A \bar{A}}}, \label{Nurowskisolnp} \\
 c(\zeta, \bar{\zeta}) = \frac{1}{A} \left[\partial_\zeta A
 + \frac{2 C_1}{\tanh \left(\frac{\rmi}{2} C_1 (z + C_0) \right)} + C_1
 \right], \label{Nurwoskisolnc} \\
 z = -\rmi \left(\int \frac{1}{A} \mathrm{d} \zeta - \int \frac{1}{\bar A} \mathrm{d}
 \bar{\zeta} \right).
\end{eqnarray}
The flexibility of choosing the function $A(\zeta)$ and real constant $C_{0,1}$
may perhaps facilitate a possible future application of the solution.
From this extended version, one can obtain the original form of Nurowski \cite{Nurwoski08} by setting
\begin{equation*}
 A(\zeta) = C_1 \zeta, \qquad \bar{A}(\bar{\zeta}) = C_1 \bar{\zeta}, \qquad C_0 = 0,
\end{equation*}
and consequently,
\begin{equation} \label{Nurowskisoln}
 p(\zeta, \bar{\zeta}) = \pm \frac{\rmi \sqrt{3}}{s (\zeta - \bar{\zeta})},
 \ c(\zeta, \bar{\zeta}) = \frac{4}{\zeta - \bar{\zeta}},
 \ \Psi_4 = \frac{14 s^2}{3 y^2} \mathrm{e}^{-\rmi r/2} \cos^3 \left(\frac{r}{2}\right).
\end{equation}
Note that all $C_1$'s are canceled out in the above expressions.
Hence another way of obtaining \eref{Nurowskisoln} is by taking the limit $C_1\rightarrow 0$
in \eref{Nurowskisolnp} and \eref{Nurwoskisolnc} (cf. \eref{JTNS}) and setting $C_0=0$ and $A(\zeta)=2$.

Modulo possible sign differences in $\alpha_I$ and $\eta_I$ caused by square roots as already discussed,
the Cartan invariants calculated from \eref{JTN} and \eref{JTNS} are both given by
\begin{eqnarray}
 \eqalign{ \alpha_I = \frac{1}{\varepsilon} \sqrt{\frac{1}{2}\sqrt{\frac{3}{5}}}, \qquad
 \beta_I = -\frac{1}{2}\sqrt{\frac{3}{5}}, \qquad
 \gamma_I = \frac{1}{2}\sqrt{\frac{3}{5}}, \cr
 \theta_I = \rmi \sqrt{\frac{3}{5}}, \qquad
 \eta_I = -\frac{1}{\varepsilon} \cdot \frac{2^{3/2}}{3^{1/4} \cdot 5^{3/4}}, \qquad
 \zeta_I = -\frac{1}{20}, \qquad \varepsilon=\pm 1.} \label{CartanInvar}
\end{eqnarray}
Like \eref{CartanInvarPFlat}, they are all constant and do not depend on $C_1$.

\section{An example of power series solutions}
For simplicity, assume that $C_1=0$ in the ODE \eref{JEQ}.
Now consider the power series solution of \eref{JEQ} satisfying the regular initial conditions
$J(0)=0$, $J'(0)=u_0>0$ and $J''(0)=0$.
A simple calculation gives us the first few terms of this series
\begin{equation} \label{JSeries}
 J(z) = \sum_{i=0}^{\infty} u_i z^{i+1} = u_0 z - \case{5}{9} \Lambda u_0^2 z^3 + \case{16}{45} \Lambda^2 u_0^3 z^5 + \cdots.
\end{equation}
Moreover, this series solution, convergent in a neighborhood of $z=0$ according to the Cauchy existence and uniqueness theorem,
is of type N with a non-vanishing Weyl scalar $\Psi_4 \propto K(z)$. Particularly,
\begin{equation*}
 K(0) = -\case{2}{3}\Lambda u_0^2 \neq 0.
\end{equation*}
To see that the solution \eref{JSeries} is not equivalent to the Leroy-Nurowski solution,
we calculate the first Cartan invariant $\alpha_I(z)$ via \eref{alphaI} which, in this case, is no longer a constant.
In particular, this series solution has
\begin{equation*}
 \alpha_I(0)=0,
\end{equation*}
with $\alpha_I(z)$ and also $K(z)$ continuous at $z=0$,
while the values given in \eref{CartanInvar} are always non-zero constants.
This is sufficient to assert
that the ODE \eref{JEQ} as well as its reductions \eref{PEQ} and \eref{Abel}
indeed contain new twisting type N solutions.

\section{One-parameter deformation from a conformally flat solution to the Leroy-Nurowski solution}

One feature that makes \eref{PEQ} preferable to the other two ODEs \eref{JEQ} and \eref{Abel} is that
the conformally flat solution \eref{PFlat1} and the extended Leroy-Nurowski solution \eref{PTN}
are just simple quadratic functions, without poles in the complex plane,
compared to their counterparts \eref{JFlatS} and \eref{JTN}.
Also note that these quadratic solutions with $C_1=0$ do not correspond to any solution of the Abel equation \eref{Abel}
since the form \eref{PTNA} with the non-constant function $Z(J)$, excludes all quadratic functions as solutions.
These well-behaved quadratic solutions facilitate a study of the power series solutions near them.
Additionally in Appendix E, we also present a Puiseux series solution to \eref{PEQ},
the existence of which is suggested by the weak Painlev\'e tests performed in Appendix D.

To simplify the notation, we apply the scaling transformation\footnote{
Once having a solution $g(w)$, one may choose a sign for $\Lambda$ in order to have $P(J)>0$.}
$J=C_1 w/\Lambda$, $P(J)=C_1^2 g(w)/\Lambda$ with $\Lambda \neq 0$, $C_1 \neq 0$
such that \eref{PEQ} takes on the form already noted as \eref{geqn} with $C=1$:
\begin{equation} \label{PEQy}
 g'' = - \frac{(g' + 2 w)^2}{2 g}  - \frac{2}{g} - \frac{10}{3}.
\end{equation}
We look for power series solutions for this equation corresponding to
the regular initial conditions $g(0)=u_0\neq 0$, $g'(0)=0$.
The first few terms of this series read
\begin{eqnarray}
 \eqalign{ \fl g(w) = \sum_{j=0}^\infty u_j w^j \cr
 \fl = u_0 -\frac{5\left(u_0+\frac{3}{5}\right)}{3u_0} w^2
 - \frac{2\left(u_0+\frac{3}{4}\right)(u_0+6)}{27u_0^3} w^4
 - \frac{76\left(u_0+\frac{3}{4}\right)(u_0+6)\left(u_0+\frac{33}{38}\right)}{1215 u_0^5} w^6 + \cdots,} \label{Powerseries}
\end{eqnarray}
where all odd order terms vanish.
The remainder of the coefficients in the series can be determined by a recursion relation
which is valid beginning with $u_6$:
\begin{equation} \label{recur}
 \fl 0 = (2k+1)(k+1) u_0 u_{2k+2} + \left(2k+\case{5}{3}\right) u_{2k}
   + \sum_{l=0}^{k-1} (k+l+1)(l+1) u_{2l+2} u_{2k-2l}, \ k\geq 2,\
\end{equation}
while $u_2$ and $u_4$ can be easily read off from \eref{Powerseries}.
It is clear that this relation allows one to calculate the coefficients to whatever order desired.
One can easily see that the coefficient of $w^{2k}$, namely $u_{2k}$,
is a $k$th-order polynomial, $P_k(u_0)$, divided by $u_0^{2k-1}\neq 0$.
Remarkably, this infinite series reduces to simple quadratic functions in two special cases.
The reason for this is that for every value of $k\geq 2$,
the polynomial $P_k(u_0)$ has the factors $\left(u_0+\frac{3}{4}\right)(u_0+6)$,
as can be seen in the few terms demonstrated in \eref{Powerseries} above and can easily be shown by induction.
Hence for $u_0=-\frac{3}{4}$, we retrieve the Leroy-Nurowski solution \eref{PTN}, which in this notation is simply
\begin{equation} \label{yLN}
 g_{LN} = -\left(\case{1}{3} w^2 + \case{3}{4}\right).
\end{equation}
As well, for $u_0=-6$, we retrieve a conformally flat solution \eref{PFlat1}, which has the form
\begin{equation} \label{yF}
 g_{CF} = - \left(\case{3}{2} w^2 + 6 \right).
\end{equation}
For all other values of $u_0\neq 0$, the formal series solution \eref{Powerseries} may then be viewed as
a generalization of these two known solutions, in terms of a power series with infinitely many terms.
It is interesting that in every one of these polynomials, $P_k(u_0)$,
all coefficients are negative, so that the only possible real roots would be negative.
Our numerical calculations suggest that none are smaller than $-6$,
and that there are no other roots common to all these different polynomials.

The series \eref{Powerseries} does define, in the complex domain, a function holomorphic in some neighborhood of the origin
as is shown by the following method of determining a non-zero radius of convergence for it.
We present the proof in Appendix F.

\textbf{Theorem 4} Given the series \eref{Powerseries} with the recursion relation \eref{recur} and a fixed $u_0\neq 0$,
one has the following bound:
\begin{equation} \label{upbd}
 |u_{2j}|\leq \frac{C M^{2j}}{(2j)^2}, \ j = 2,3,\cdots,
\end{equation}
provided that one can pick two constants $C>0$ and $M>0$ such that they satisfy
\begin{eqnarray}
 && \left|\frac{2\left(u_0+\frac{3}{4}\right)(u_0+6)}{27u_0^3} \right|
 \leq \frac{C M^4}{16}, \label{ineq1} \\
 && \left(\frac{5}{3} + \frac{1}{|u_0|}\right) \frac{9}{4 M^2}
 + \left(\frac{\pi^2}{12} - \frac{1}{4}\right) C \leq |u_0|. \label{ineq2}
\end{eqnarray}

The existence of such an upper bound \eref{upbd} on $u_{2j}$
guarantees a lower bound $M^{-1}$ on the radius of convergence.
For instance, if we take $u_0 = -2$, which lies nicely in the interval between $-\frac{3}{4}$ and $-6$,
we can at least pick
\begin{equation*}
 C=\case{1}{10}, \qquad M^{-1}=\case{3}{5}
\end{equation*}
satisfying both \eref{ineq1} and \eref{ineq2}.
The bound \eref{upbd} is by no means optimal at every $u_0\neq 0$.
In fact, our numerical integrations of \eref{PEQy} with $u_0$ sampled between $-6$ and $-\frac{3}{4}$
all indicate that in the \emph{real} domain,
the series solutions \eref{Powerseries} with $-6 < u_0 < -\frac{3}{4}$
are all well sandwiched between the parabolic curves of \eref{yLN} and \eref{yF},
and therefore suggest an infinite radius of convergence on the real line.
Moreover, by applying the transformation $w \rightarrow \frac{1}{w}$ to \eref{PEQy}
and studying the formal (Puiseux) series expansion of the transformed ODE at the origin,
we find the following asymptotic expansion\footnote{
We also find another asymptotic expansion that has the first two leading terms identical to \eref{yLN},
but also involves fractional powers of $w$ in a complicated way, hence not presented here.}
of \eref{PEQy} as $w\rightarrow \infty$ (cf. \eref{yF}):
\begin{eqnarray*}
  g \sim -\case{3}{2} w^2 - 6 + u_{4/3} w^{2/3} + O(w^{-1/3}),
\end{eqnarray*}
where $u_{4/3}$ is an arbitrary constant.
This asymptotic behaviour at infinity, consistent with our numerical calculations, again suggests that
we may significantly extend the radius of convergence for \eref{Powerseries} at least in the real domain.

An additional comment is that the Cartan invariant $\alpha_I$, computed from \eref{Powerseries} with $-6<u_0<-\frac{3}{4}$,
is not constant and has a dependence on $C_1$, contrary to the special cases for those values of $u_0$
at the two endpoints of the interval of values for $u_0$ being considered.

\section{Conclusions}

We have begun with the advantage of prior work done on the use of (3-dimensional) CR manifolds
to look for solutions of the Einstein field equations that correspond to algebraically-degenerate Einstein spaces,
with twisting principal null directions.
A general solution of those reduced field equations for the two functions of three variables
would generate all twisting solutions of Petrov type N.
Of course we did not achieve this;
however, after the assumption of a single Killing vector in a particular direction,
our ansatz for group-invariant solutions obtained from the infinite-dimensional classical symmetries of the field equations,
allowed us to obtain a single ODE, the solutions of which would generate a family of solutions of type N
with twisting principal null directions.
That ODE is either a rather simple, third-order nonlinear equation for $J=J(z)$
in which the independent variable $z$ does not appear or,
equivalently, an even simpler, second-order nonlinear equation for $g=g(w)$,
where $w$ is a dimensionless re-scaling of $J$ and $g$ is a re-scaling of $J'$,
which includes a non-zero value for $\Lambda$, the cosmological constant.
Within the same ansatz, we have also investigated all the cases of solutions corresponding to conformally flat spacetimes
to which type N solutions may degenerate, which helps us look for non-trivial cases.

We have studied this second-order equation at some length.
In particular it contains one parameter, $C_1$, which may always be re-scaled to the value $+1$ unless it happens to be zero.
In the case that it is zero, the equation can be reduced still further, to a first-order equation of Abel type.
Following standard approaches to Abel equations we were unable to determine any method
that we thought would generate reasonable type N solutions, although this is still an ongoing project of considerable interest.
However, when $C_1$ is not zero we have considered various sorts of solutions which it might have.
We have shown that it does have solutions which are holomorphic, in the complex plane in a neighborhood of the origin,
and have found an asymptotic behavior near the (real) infinity.
In particular we have picked out especially those solutions which are even functions of $w$
and looked at power-series solutions about the origin, both analytically and numerically via a Maple computer program.
We have determined a moderately-simple recursion relation for the coefficients of the powers of $w^2$ in the series solutions,
which determines the coefficient $u_{2k+2}$, of $w^{2k+2}$ ($k\geq 2$) in terms of all the previous coefficients,
looking at all of them as determined by the value of $g(0)=u_0$.
This series terminates quickly for just two particular values of $u_0$, in the form $-a(u_0)w^2+u_0$,
with $a$ constant, different for the two values of $u_0$.
The value $u_0=-\frac{3}{4}$ generates the previously-known Leroy-Nurowski solution,
while the other one $u_0=-6$ is unfortunately simply a conformally flat solution.
To ensure that these series solutions are distinct from the Leroy-Nurowski solution,
we have used the work of Cartan on the question of the equivalence of two CR manifolds,
which requires the equality of the set of six Cartan invariants.
We have found that any value of $u_0$ between these two special values generates Cartan invariants
that are quite different from those at the endpoints of this interval,
and therefore distinct from those of the Leroy-Nurowski solution.

The solutions characterized by values of $u_0$ between $-6$ and $-\frac{3}{4}$ have an asymptotic behavior,
via a Puiseux series around the (real) infinity,
that has the same form $-a(-6) w^2-6$ as the conformally flat solution aforementioned,
but also lower-order terms involving third-roots of $w$,
which doubtless generate algebraic singularities there.
Numerical integrations via Maple agree with this behavior,
showing negative values of $g(w)$ as needed and very simple structure, for all real values of $w$.
The same numerical integrations do show singularities in the solutions for $u_0>-\frac{3}{4}$.
As well, numerical calculations of the coefficients $u_{2k+2}$,
for several values of $u_0\in \left(-6,-\frac{3}{4}\right)$ (e.g., $u_0=-\frac{301}{400}$)
show that starting at a large enough $k$, they alternate in sign
while their absolute values are monotonically decreasing at rapid rates.
We therefore postulate that these solutions are everywhere non-singular and well-behaved on the real axis,
and believe that they might define new well-behaved, transcendental functions with algebraic singularities off the real $w$-axis.
The proof of such a conjecture is still being pursued;
nonetheless, we feel that the numerical calculations justify the belief
that this is a sufficiently interesting result as to merit the attention of a wider audience.

To conclude the discussion, we present here our new class of metrics,
which, without loss of generality, may be considered by setting $A(\zeta)=2$ in the ansatz \eref{pccb}.
Although we will present it here with the new real coordinate $z$ introduced in \eref{zdef},
with this choice of $A(\zeta)$ it is the same as the usual coordinate $y$ used in \eref{lambda}.
As well, our studies with the equation for $P(J)$, equivalently $g(w)$,
allow us to replace $z$ by its form in terms of $J$ as determining the imaginary part of $\rmd \zeta$,
via $\rmd z = \rmd J/P(J)$, namely,
\begin{equation*}
 \zeta = x+\rmi z = x+\rmi z(J),
 \qquad \rmd\zeta = \rmd x + \rmi \rmd z  = \rmd x + \frac{\rmi}{P} \rmd J.
\end{equation*}
For simplicity of presentation, we show both forms below,
with coordinates $\{x,z,u,r\}$ or $\{x,J,u,r\}$:
\begin{equation*}
 \mathbf g = \frac{J'}{2 \cos^2(\frac{r}{2})} \left[ \rmd\zeta \rmd\bar\zeta + \lambda \left(\rmd r + W \rmd\zeta + \bar W \rmd\bar\zeta
 + H \lambda \right) \right]
\end{equation*}
with real-valued $J=J(z)$, $J'\equiv \rmd J/\rmd z =P(J)>0$ and $P' \equiv \rmd P/\rmd J$ such that
\begin{eqnarray*}
 W = \frac{1}{2} \left( \frac{J''}{2 J'} + \Lambda J + \rmi C_1 \right) (\mathrm{e}^{-\rmi r}+1)=
 \case{1}{2} \left( \case{1}{2} P' + \Lambda J + \rmi C_1 \right) (\mathrm{e}^{-\rmi r}+1), \\
 H = -\case{1}{6} \Lambda J' \cos(r) = -\case{1}{6} \Lambda P \cos(r).
\end{eqnarray*}
where $C_1$ is an arbitrary real parameter.
The function $L$ as in $\partial=\partial_\zeta - L \partial_u$ can be chosen so as to be real-valued:
\begin{equation*}
 L = -\mathrm{e}^{-C_1 x} \int \exp\left(\int F_2 \mathrm{d}z \right) \mathrm{d}z
 = -\mathrm{e}^{-C_1 x} \int \frac{1}{P} \exp\left(\int \frac{F_2}{P}\, \rmd J \right) \rmd J,
 \end{equation*}
such that from \eref{lambda},
 \begin{equation*}
 \fl \lambda = \frac{\mathrm{e}^{C_1 x} \rmd u
 - 2 \left[\int \exp\left(\int F_2 \mathrm{d}z \right) \mathrm{d}z \right] \rmd x}
 {\exp\left(\int F_2 \mathrm{d}z \right)}
 = \frac{\mathrm{e}^{C_1 x} \rmd u
 - 2 \left[\int P^{-1} \exp\left(\int F_2 P^{-1} \rmd J \right) \rmd J \right] \rmd x}
 {\exp\left(\int F_2 P^{-1} \rmd J \right)},
\end{equation*}
where $F_2$ is given by
\begin{equation*}
 F_2 = \frac{J^{\prime\prime}}{2 J'} - \Lambda J = \case{1}{2} P' - \Lambda J.
\end{equation*}
Meanwhile, the functions $J(z)$ and $P(J)$ respectively satisfy
\begin{eqnarray*}
 J''' = \frac{(J^{\prime\prime})^2}{2J'} - 2 \Lambda J J^{\prime\prime} -
 \frac{10}{3} \Lambda (J')^2 - 2 (\Lambda^2 J^2 + C_1^2) J',\\
 P'' = - \frac{(P' + 2 \Lambda J)^2}{2P} - \frac{2 C_1^2}{P} - \frac{10}{3} \Lambda.
\end{eqnarray*}
In particular, the original metric by Nurowski \cite{Nurwoski08} corresponds to the case
$C_1=0$, $J=\frac{3}{\Lambda z}$ and a proper choice of the integration constants in $\lambda$.

We can here note the philosophy that certain ODEs themselves may serve the purpose of defining \emph{new} transcendental functions;
for instance, we recall the Painlev\'e functions and the associated ODEs.
Hence our situation with new type N solutions being determined by a second-order \emph{nonlinear} ODE
is presumably not too different from that of Hauser's solution (in terms of hypergeometric functions \cite{Hauser74})
which is determined by a second-order \emph{linear} ODE,
although it is true that there has already been much more extensive studies made on the properties of hypergeometric functions
than have been made for newer functions defined by solutions of nonlinear ODEs
that may not even have the Painlev\'e property.

\appendix
\section{The Abel ODE}

The equation \eref{Abel} actually does have the following special solution
\begin{equation} \label{PFlat3}
 f_{CF} = -\frac{3}{4t+6}.
\end{equation}
However, it can be shown to correspond to a conformally flat solution \eref{PFlat2},
and hence is not interesting.

Unfortunately, we have had no luck so far finding the general solution to \eref{Abel}
or any other special solution other than \eref{PFlat3}.
Since constructing the general solution to the generic Abel ODE has remained an open problem for decades,
the general strategy of integration nowadays mainly lies in recognizing, within a suitable class of transformations,
the ODE in question as equivalent to a previously solved equation.
Such a procedure has been programmed into the current state-of-the-art Maple code \texttt{dsolve} (or \texttt{abelsol}) \cite{Cheb-Terrab00,Cheb-Terrab03},
which presumably covers all/most of the integrable classes presented in Kamke's book \cite{Kamke59}
and various other references (e.g., \cite{Polyanin95}).
However, this code, as tested by us, does not recognize \eref{Abel} as a known solved type, e.g., the AIR class.
Other attempts by us, such as the symmetry method, on finding special solutions all have failed or just led to \eref{PFlat3}.

So far we have not been able to find a similar reduction for the ODE \eref{PEQ} with $C_1\neq 0$,
nor can we negate the possibility that \eref{PEQ} with $C_1\neq 0$
may contain different type N solutions other than the case with $C_1=0$.
In fact, the Cartan invariants calculated with \eref{JEQ} generally do have a dependence on the constant $C_1$
even though this is not the case for all the conformally flat solutions and the Leroy-Nurowski solution
(see \eref{CartanInvarPFlat} and \eref{CartanInvar}).

\section{Cartan invariants}

A concise description of Cartan invariants can be found in Section 2.1.3 of \cite{Nurowski93}.
Given $c=c(\zeta,\bar\zeta)$ and
\begin{equation*}
 r = \case{1}{6}\left(\partial_{\bar\zeta} \bar{l} + 2 \bar{c} \bar{l}\right), \qquad
 l = -\partial_\zeta \partial_{\bar\zeta} c - c \partial_{\bar\zeta} c
\end{equation*}
taken from \eref{alphaI} where $\alpha_I$ is presented,
the next three Cartan invariants, when $r\neq 0$, read
\begin{eqnarray*}
 \beta_I(\zeta,\bar\zeta) =& \frac{1}{32 (r \bar{r})^{9/4}}
 \Big[ 3 \bar{r}^2 \partial_{\bar\zeta} r\, \partial_\zeta r
 + 3 r^2 \partial_\zeta \bar{r}\, \partial_{\bar\zeta} \bar{r}
 - r\bar{r} \Big( \partial_\zeta \bar{r}\, \partial_{\bar\zeta} r
 \\
 & + 7 \partial_\zeta r\, \partial_{\bar\zeta} \bar{r}
 + 16 \bar{c} \bar{r} \partial_\zeta r + 16 c r \partial_{\bar\zeta} \bar{r}
 - 8 r \bar{r} \partial_{\bar\zeta} c + 16 c \bar{c} r \bar{r} \Big) \Big],
\end{eqnarray*}
\begin{eqnarray*}
 \gamma_I(\zeta,\bar\zeta) =& \frac{-1}{32 (r \bar{r})^{9/4}}
 \Big[7 \bar{r}^2 \partial_{\bar\zeta} r\, \partial_\zeta r
 + 7 r^2 \partial_{\bar\zeta} \bar{r}\, \partial_\zeta \bar{r}
 - r\bar{r} \Big( 8 r \partial_\zeta \partial_{\bar\zeta} \bar{r}
 + 8 \bar{r} \partial_\zeta \partial_{\bar\zeta} r
 \\
 & + \partial_\zeta \bar{r}\, \partial_{\bar\zeta} r
 + \partial_\zeta r\, \partial_{\bar\zeta} \bar{r}
 + 4 c \bar{r} \partial_{\bar\zeta} r + 4 \bar{c} r \partial_\zeta \bar{r}
 + 4 c r \partial_{\bar\zeta} \bar{r} + 4 \bar{c} \bar{r} \partial_\zeta r
 \\
 & + 24 r \bar r \partial_{\bar\zeta} c + 16 c \bar{c} r \bar{r} \Big)\Big],
\end{eqnarray*}
\begin{eqnarray*}
 \theta_I(\zeta,\bar\zeta) = & \frac{-\rmi}{16 r (r \bar{r})^{7/4}}
 \Big[ 5 \bar{r}^2 (\partial_{\bar\zeta} r)^2
 + 5 r^2 (\partial_{\bar\zeta} \bar{r})^2
 - r\bar{r} \Big( 4 r \partial_{\bar\zeta}^2 \bar{r}
 + 4 \bar{r} \partial_{\bar\zeta}^2 r
 \\
 & - 2 \partial_{\bar\zeta} r \partial_{\bar\zeta} \bar{r}
 - 4 \bar{c} \bar{r} \partial_{\bar\zeta} r - 4 \bar{c} r \partial_{\bar\zeta} \bar{r}
 + 16 r \bar{r} \partial_{\bar\zeta} \bar{c} \Big)\Big].
\end{eqnarray*}
Due to the length of $\zeta_I$ and $\eta_I$ as calculated with Maple for our studies, we will not put them here.
All Cartan invariants are uniquely determined by the function $c=c(\zeta,\bar\zeta)$.

\section{Conformally flat solutions}

For a further integration of \eref{JFlatG}, we have three separate cases.

\emph{Case 1:} $\Lambda<0,\ C_2>0$. We always have $J'\geq 0$. Then the solution is determined by
\begin{eqnarray*}
 \eqalign{ \ln \frac{G^2+\sqrt{2}GM+M^2}{G^2-\sqrt{2}GM+M^2}
 + 2\arctan\left(\frac{\sqrt{2}GM}{M^2-G^2}\right) = -2\sqrt{2} M^3(z+C_0), \cr
 M=\pm\left(-\frac{2C_2 \Lambda^{1/3}}{3}\right)^{1/4},\qquad G=(\Lambda J)^{1/3}.}
\end{eqnarray*}
In the real domain, the inverse function $J=J(z)$ is well defined over
$z+C_0 \in \left(-\frac{\pi}{\sqrt{2}|M^3|}, \frac{\pi}{\sqrt{2}|M^3|} \right)$
instead of the entire real line, and has singularities at $z+C_0=\pm \frac{\pi}{\sqrt{2}|M^3|}$.

\emph{Case 2:} $\Lambda<0,\ C_2<0$. We need $|J|\geq \left(2C_2/3\Lambda\right)^{3/4}$ for $J'\geq 0$.
The solution is determined by
\begin{eqnarray*}
 \eqalign{ \ln\left|\frac{M+G}{M-G}\right| + 2\arctan\left(\frac{G}{M}\right)
 = 2 M^3(z+C_0), \cr
 M=\pm\left(\frac{2C_2 \Lambda^{1/3}}{3}\right)^{1/4},\qquad G=(\Lambda J)^{1/3},
 \qquad |J|\geq \left(\frac{2C_2}{3\Lambda}\right)^{3/4}.}
\end{eqnarray*}
In the real domain, the inverse function $J=J(z)$ is well defined over
$z+C_0 \in \left(-\infty, -\frac{\pi}{2|M^3|}\right) \cup \left(\frac{\pi}{2|M^3|}, +\infty \right)$,
and has singularities at $z+C_0=\pm \frac{\pi}{2|M^3|}$.

\emph{Case 3:} $\Lambda>0,\ C_2>0$. We need $|J|\leq \left(2C_2/3\Lambda\right)^{3/4}$ for $J'\geq 0$.
The solution is determined by
\begin{eqnarray*}
 \eqalign{ \ln\left|\frac{M+G}{M-G}\right| + 2\arctan\left(\frac{G}{M}\right)
 = 2 M^3(z+C_0), \cr
 M=\pm\left(\frac{2C_2 \Lambda^{1/3}}{3}\right)^{1/4},\qquad G=(\Lambda J)^{1/3},
 \qquad |J|\leq \left(\frac{2C_2}{3\Lambda}\right)^{3/4}.}
\end{eqnarray*}
In the real domain, the inverse function $J=J(z)$ from above is well defined over the entire real line.

\section{Weak Painlev\'{e} tests}

We have three ODEs \eref{JEQ}, \eref{PEQ}, and \eref{Abel} at hand
that may be explored for new twisting type N solutions.
A particular, probably useful way to decide which one of these equations has a better chance for one to find a solution
is given by the (weak) Painlev\'{e} test \cite{Conte99,Conte08,Ramani89}.
This test reveals the nature of the movable singularities of the general solution of a nonlinear ODE.
Failing the test means the occurrence of certain undesirable movable singularities,
e.g., infinitely branched singularities, that relate to non-integrability \cite{Ramani89},
although it may still be possible to find special solutions.
Associated to the (weak) Painlev\'{e} test is the global property called the (weak) Painlev\'{e} property.
An ODE possesses the Painlev\'{e} property if the general solution can be made single-valued.
For the weak Painlev\'{e} property, it requires that the general solution be at most finitely branched around any movable singularity.
The tests themselves are by design sets of necessary conditions respectively for these properties.

In this appendix, we will show that none of the three ODEs pass the Painlev\'{e} test,
and that \eref{JEQ} also fails the weak Painlev\'{e} test while the other two pass.
To begin, we detail the test procedures on \eref{PEQ}.
Then we briefly comment on \eref{Abel}
and simply point out where the tests fail for \eref{JEQ} without dwelling on details.

The equation \eref{PEQ} surely does not have the Painlev\'{e} property
for the coefficient of the $(P')^2$ term clearly violates the necessary conditions
for the Painlev\'{e} property \cite{Conte99} (see p.127).
This is also confirmed by the test conclusion that \eref{PEQ} has movable algebraic singularities.

\textit{Step 1} (\textit{Dominant behaviours}). Assume the leading behaviour of a solution $P(J)$ to be
\begin{equation*}
  P \sim u_0 \chi^m, \qquad \chi=J-J_0, \qquad u_0\neq 0, \qquad m\neq 0,
\end{equation*}
with $m$ not a positive integer.
Substitute this form into \eref{PEQ} and select out all possible lowest order terms as listed below
\begin{equation*}
  \case{3}{2} u_0^2 m \left(m - \case{2}{3} \right) \chi^{2m-2}, \qquad
  2\Lambda u_0 J_0 m \chi^{m-1}, \qquad
  2(\Lambda^2 J_0^2 + C_1^2).
\end{equation*}
Since $m\neq 1$, we only have two possibilities.
For $m<1$, $\chi^{2m-2}$ is the lowest order term
and the vanishing of its coefficient requires
\begin{equation*}
  m = \frac{2}{3}
\end{equation*}
given $u_0, m\neq 0$.
For $m>1$, the constant $2(\Lambda^2 J_0^2 + C_1^2)$ is the lowest order term,
which does not vanish in general, hence not interesting for the purpose.
To summarize, we obtain $m=\frac{2}{3}$ with arbitrary $u_0\neq 0$, i.e, that
\begin{equation*}
 P \sim u_0 (J-J_0)^{2/3}
\end{equation*}
is the only detected dominant behaviour.

\textit{Step 2} (\textit{Resonance conditions} \cite{Conte99} (see p.87)).
Having found the dominant behaviour,
now we consider the possibility to extend it to a Puiseux series expansion
\begin{equation*}
  P = \sum_{j=0}^\infty u_j (J-J_0)^{(j+2)/3}.
\end{equation*}
This requires the determination of the locations $(j+2)/3$, called \emph{Fuchs indices} or \emph{resonances},
where arbitrary coefficients may enter the Puiseux series.
Consider the dominant terms
\begin{equation*}
 \hat{E}(J,P) = P P'' + \case{1}{2} (P')^2
\end{equation*}
of \eref{PEQ} that contribute to the leading behaviour $\chi^{2m-2}=\chi^{-2/3}$.
Then compute the derivative
\begin{equation*}
  \lim_{\epsilon\rightarrow 0} \frac{\hat{E}(J,P+\epsilon V) - \hat{E}(J,P)}{\epsilon}
  = (P \partial^2_J + P' \partial_J + P'') V.
\end{equation*}
The Fuchs indices satisfy the so-called \emph{indicial equation}
\begin{eqnarray*}
  \lim_{\chi\rightarrow 0} \chi^{-j-(2m-2)}
  (P \partial^2_J + P' \partial_J + P'') \chi^{j+m}
  = u_0 (j+1) j = 0.
\end{eqnarray*}
Hence we obtain a fractional resonance at $(j+2)/3=\frac{2}{3}$ with $j=0$.

\textit{Step 3 (Compatibility conditions).}
At $j=0$, we know, from the the first step, that $u_0(\neq 0)$ is indeed an arbitrary coefficient.
This completes the test.
In conclusion, \eref{PEQ} passes the weak Painlev\'{e} test.

\textbf{Remark} Note that no \emph{pole} is detected from the test above.
The ODE for $P^3$ still involves a Puisuex series instead of a Laurent series
since the cubing does not eliminate all third roots of $\chi$.
According to \cite{Ramani89},
The presence of movable algebraic singularities is not incompatible with integrability.

The very design of the weak Painlev\'{e} test limits its usage only as necessary conditions for the weak Painlev\'{e} property.
The test can neither detect movable (branched) essential singularities themselves
nor exclude an accumulation of algebraic singularities forming a movable essential one that may be severely branched.
These possibilities make a rigorous proof of the weak Painlev\'{e} property not at all a trivial one,
which by itself may deserve a specialized article to discuss.
See examples in \cite{Smith53,Shimomura05,Filipuk09}.

According to Painlev\'{e} \cite{Painleve88,Hille76},
the only movable singularities of solutions to the first-order ODE $y'=F(x,y)$
where $F$ is rational in $y$ with coefficients that are algebraic functions of $x$,
are poles and/or algebraic branch points.
In addition, the only nonlinear ODE in this class that has the Painlev\'{e} property
is the Riccati equation which \eref{Abel} is certainly not.
Hence the equation \eref{Abel} automatically has the weak Painlev\'{e} property, but not the Painlev\'{e} property,
and it is free from movable essential singularities.

The equation \eref{JEQ} admits two families of dominant behaviours (cf. \eref{JFlatSS} and \eref{JTNS}):
\begin{eqnarray*}
 J \sim \frac{2}{3\Lambda (z-z_0)}, \qquad \textrm{Fuchs indices}=-1,\frac{4}{3},\frac{7}{3}; \\
 J \sim \frac{3}{\Lambda (z-z_0)}, \qquad \textrm{Fuchs indices}=-1,-\frac{1+\sqrt{57}}{2},-\frac{1-\sqrt{57}}{2}.
\end{eqnarray*}
It fails the weak Painlev\'{e} test for having \emph{irrational} resonances.
This means that \eref{JEQ} has an infinitely branched movable singularity,
which is a strong indicator for non-integrability \cite{Ramani89}.

Since our attempt of solving \eref{Abel} has not been successful,
we decided to focus on \eref{PEQ} and explore some of its features that may facilitate constructing new solutions.

\section{Puiseux series solutions}

As indicated by the weak Painlev\'{e} test, the ODE \eref{PEQ} for $P(J)$
possesses a formal Puiseux series solution
\begin{eqnarray} \label{Puiseux}
 \fl P = \sum_{k=0}^\infty u_k (J-J_0)^{(k+2)/3} \nonumber \\
 \fl   = u_0 (J-J_0)^{2/3} - 3 \Lambda J_0 (J-J_0) - \frac{9(\Lambda^2 J_0^2+4C_1^2)}{20u_0} (J-J_0)^{4/3} \nonumber \\
 \fl  - \frac{3\Lambda J_0 (\Lambda^2 J_0^2+4C_1^2)}{5u_0^2} (J-J_0)^{5/3}
   - \left[\frac{3}{2}\Lambda + \frac{27(109\Lambda^2 J_0^2+36C_1^2)(\Lambda^2 J_0^2+4C_1^2)}{2800 u_0^3}\right] (J-J_0)^2 + \cdots
   \nonumber \\
\end{eqnarray}
with two arbitrary complex constants $u_0\neq 0$ and $J_0$.
In particular, this Puiseux series solution contains a special case for $J_0=\pm 2\rmi C_1/\Lambda$ ($\Lambda\neq 0$) such that
\begin{equation*}
 P = u_0 \left(J \pm \frac{2\rmi C_1}{\Lambda}\right)^{2/3} \!\!\!\!
 - \frac{3}{2} \Lambda \left(J^2 + \frac{4C_1^2}{\Lambda^2} \right).
\end{equation*}
This finite expression coincides with the known solution \eref{PFlat} (setting $u_0=C_2$).

\textbf{Theorem 5} Given that $u_0\neq 0$ and $u_0$, $J_0\in \mathbb{C}$,
the ODE \eref{PEQ} admits a formal Puiseux series solution \eref{Puiseux}
such that it converges in a neighborhood of $J_0$.

\emph{Proof:} The idea of the proof, following many standard proofs of Painleve property,
is to convert the Puiseux series into a power series solution of a regular initial value problem (e.g., \cite{Smith53,Filipuk09}).
First we define
\begin{equation} \label{ZEQ1}
 Z = P^{1/2} (P' + 4 \Lambda J).
\end{equation}
Then differentiate it once with respect to $J$ and substitute $P''$ using \eref{PEQ}. Hence we obtain
\begin{equation} \label{ZEQ2}
  Z' = \frac{2(\Lambda P - 3 \Lambda^2 J^2 - 3 C_1^2)}{3 P^{1/2}}.
\end{equation}
The system (\ref{ZEQ1},\ref{ZEQ2}) is equivalent to the ODE \eref{PEQ}.
Now by introducing a new variable $U=P^{1/2}$, we can transform the system into
\begin{eqnarray}
 \eqalign{ \frac{d J}{d U} = \frac{2U^2}{Z-4\Lambda U J}, \cr
 \frac{d Z}{d U} = -\frac{4(3\Lambda^2 J^2 - \Lambda U^2 + 3 C_1^2) U}{3(Z-4\Lambda U J)}.} \label{JZUEQ}
\end{eqnarray}
which has a unique power series solution about $U=0$
\begin{eqnarray}
 \eqalign{ J = J_0 + \frac{2}{3Z_0} U^3 + \cdots, \cr
 Z = Z_0 - \frac{2(\Lambda^2 J_0^2 + C_1^2)}{Z_0} U^2 + \cdots.} \label{JZUSeries}
\end{eqnarray}
By the Cauchy existence and uniqueness theorem, both series have non-vanishing radii of convergence.
From the series \eref{JZUSeries}, the corresponding solutions to (\ref{ZEQ1},\ref{ZEQ2}) then take the form
\begin{eqnarray*}
 P = \left[\frac{3 Z_0}{2}(J-J_0)\right]^{2/3} \!\!\!\! + \sum_{k=1}^\infty u_k (J-J_0)^{(k+2)/3}, \nonumber \\
 Z = Z_0 + \sum_{k=0}^\infty v_k (J-J_0)^{(k+2)/3}.
\end{eqnarray*}
with $Z_0\neq 0$. This completes the proof.

The series \eref{Puiseux} clearly contains type N solutions that are not equivalent to Leroy-Nurowski's
since they all continuously deform to the conformally flat solution \eref{PFlat2} in the limit $J_0\rightarrow 0$, $C_1\rightarrow 0$.
We already know that the latter has a non-constant Cartan invariant $\alpha_I$ given by \eref{CartanInvarPFlat2}.

\section{Proof of Theorem 4}

The induction begins with
\begin{equation*}
 |u_4| \leq \frac{C M^4}{16}.
\end{equation*}
which holds by the assumption \eref{ineq1}.
Now assume that for $k\geq 2$ and $j=2,\cdots,k$, the bound \eref{upbd} is true.
Then for $k\geq 3$ and $1\leq l\leq k-2$, we can bound the product $u_{2l+2} u_{2k-2l}$ by
\begin{eqnarray}
 |u_{2l+2} u_{2k-2l}|
 &\leq& \frac{C^2 M^{2k+2}}{(2l+2)^2 (2k-2l)^2} \nonumber \\
 &\leq& 2\left[\frac{(2k-2l)^2+(2l+2)^2}{(2k+2)^2}\right] \frac{C^2 M^{2k+2}}{(2l+2)^2 (2k-2l)^2} \nonumber \\
 &=& 2\left[\frac{1}{(2l+2)^2}+\frac{1}{(2k-2l)^2}\right] \frac{C^2 M^{2k+2}}{(2k+2)^2}.  \label{ind2}
\end{eqnarray}
The second inequality above is due to $(a^2+b^2)/(a+b)^2\geq \frac{1}{2}$.
Rearranging \eref{recur} and using the triangular inequality together with \eref{ind2},
we obtain an upper bound for $|u_{2k+2}|$:
\begin{eqnarray} \label{tri-ineq}
 \fl |u_{2k+2}|
 &\leq& \frac{\left(2k+\frac{5}{3}\right) |u_{2k}| + (2k^2+k+1) |u_2 u_{2k}|}{(2k+1)(k+1) |u_0|}
 + \frac{\sum_{l=1}^{k-2} (k+l+1)(l+1) |u_{2l+2} u_{2k-2l}|}{(2k+1)(k+1) |u_0|} \nonumber \\
 \fl &\leq& \frac{\left(2k+\frac{5}{3}\right) + (2k^2+k+1) |u_2|}{(2k+1)(k+1) |u_0|}
  \cdot \frac{C M^{2k}}{(2k)^2}
 + \frac{S(k)}{2(2k+1)(k+1) |u_0|}\cdot \frac{C^2 M^{2k+2}}{(2k+2)^2},\ k\geq 2,
 \nonumber \\ \fl
\end{eqnarray}
where we define
\begin{equation*}
 \fl S(k) = \sum_{l=1}^{k-2} \frac{(k+l+1)(l+1)}{(l+1)^2}
 + \sum_{l=1}^{k-2} \frac{(k+l+1)(l+1)}{(k-l)^2}, \ k\geq 3,\ \textrm{and}\ S(2)=0
\end{equation*}
We can evaluate the first summation above in terms of the digamma function
\begin{equation*}
 \sum_{l=1}^{k-2} \frac{(k+l+1)(l+1)}{(l+1)^2}
 = k\Psi(k)-(2-\gamma)k \leq k\Psi(k)
\end{equation*}
where $\gamma$ is Euler's constant, which is approximately $0.57721\cdots$.
The second summation has the following bound
\begin{eqnarray*}
 \fl \sum_{l=1}^{k-2} \frac{(k+l+1)(l+1)}{(k-l)^2}
  &=& \sum_{l=1}^{k-2} \frac{(2k-l)(k-l)}{(l+1)^2} \nonumber \\
 \fl &\leq& 2k^2 \sum_{l=1}^{k-2} \frac{1}{(l+1)^2}
  = \left(\frac{\pi^2}{3} - 2\right) k^2 - 2 k^2\Psi(1,k)
  \leq \left(\frac{\pi^2}{3} - 2\right) k^2.
\end{eqnarray*}
Note that the trigamma function $\Psi(1,k)\geq 0$ for all integers $k\geq 3$
and that $\Psi(1,k)\sim k^{-1}$ for $k\rightarrow +\infty$.
Combining these two bounds, for $k\geq 3$, we obtain
\begin{eqnarray*}
 \frac{S(k)}{2(2k+1)(k+1)}
 &\leq& \frac{\left(\pi^2/3 - 2\right) k^2 + k\Psi(k)}{2(2k+1)(k+1)} \nonumber \\
 &\leq& \frac{\left(\pi^2/3 - 2\right) k^2 + k^2}{4k^2}
 = \frac{\pi^2}{12} - \frac{1}{4},
\end{eqnarray*}
where we use the fact that $0 \leq \Psi(k)\leq k$ for all integers $k\geq 3$.
In addition, the first term in \eref{tri-ineq} is bounded by
\begin{eqnarray*}
 \fl \frac{\left(2k+\frac{5}{3}\right) + (2k^2+k+1) |u_2|}{(2k+1)(k+1) |u_0|}
  \cdot \frac{C M^{2k}}{(2k)^2}
 &\leq& \frac{\left(2k+\frac{5}{3}\right) + (2k^2+k+1) \left(\frac{5}{3}+\frac{1}{|u_0|}\right)}
 {(2k+1)(k+1) |u_0|}\cdot \frac{C M^{2k}}{(2k)^2}
 \nonumber \\
 \fl &\leq& \frac{1}{|u_0|}\left(\frac{5}{3} + \frac{1}{|u_0|}\right) \frac{(k+1)^2}{k^2 M^2}\cdot
 \frac{C M^{2k+2}}{(2k+2)^2}
 \nonumber \\
 \fl &\leq& \frac{1}{|u_0|} \left(\frac{5}{3} + \frac{1}{|u_0|}\right)
 \frac{9}{4 M^2}\cdot \frac{C M^{2k+2}}{(2k+2)^2},
\end{eqnarray*}
where the last inequality becomes an equality for $k=2$.
Altogether, we obtain for $k\geq 2$
\begin{equation*}
 \fl |u_{2k+2}| \leq \frac{1}{|u_0|}
 \left[\left(\frac{5}{3} + \frac{1}{|u_0|}\right) \frac{9}{4 M^2}
 + \delta_{k}^2 \left(\frac{\pi^2}{12} - \frac{1}{4}\right) C \right]
 \frac{C M^{2k+2}}{(2k+2)^2} \leq \frac{C M^{2k+2}}{(2k+2)^2},
\end{equation*}
given the assumption \eref{ineq2}. Here $\delta_k^j$ is the Kronecker delta.
This completes the induction.


\end{document}